\documentstyle[epsfig]{mn}
\input{epsf}

\newcommand{\pl}{\partial}
\renewcommand{\d}{{\rm d}}
\newcommand{\beq}{\begin{equation}}
\newcommand{\eeq}{\end{equation}}
\newcommand{\beqa}{\begin{eqnarray}} 
\newcommand{\eeqa}{\end{eqnarray}}
\newcommand{\bea}{\begin{array}} 
\newcommand{\ea}{\end{array}} 
\newcommand{\lag}{\langle}
\newcommand{\rag}{\rangle}
\newcommand{\Om}{\Omega_{\rm m}}
\newcommand{\Ol}{\Omega_{\Lambda}}
\newcommand{\De}{{\cal D}}
\newcommand{\gam}{\gamma}
\newcommand{\cP}{{\cal P}}
\newcommand{\kappamin}{\kappa_{\rm min}}
\newcommand{\kappah}{\hat{\kappa}}
\newcommand{\wh}{\hat{w}}
\newcommand{\kaphs}{\hat{\kappa}_s}
\newcommand{\Map}{M_{\rm ap}}
\newcommand{\tS}{\tilde{S}}
\newcommand{\bx}{{\bf x}}
\newcommand{\bk}{{\bf k}}
\newcommand{\kpar}{k_{\parallel}}
\newcommand{\kperp}{\bk_{\perp}}
\newcommand{\kperpDt}{k_{\perp}\De\theta_s}
\newcommand{\gamh}{\hat{\gamma}}
\newcommand{\gamhs}{\hat{\gamma}_s}
\newcommand{\inta}{\int_{-i\infty}^{+i\infty}}
\newcommand{\rhob}{\overline{\rho}}
\newcommand{\xib}{\overline{\xi}}
\newcommand{\phikap}{\varphi_{\kaphs}}
\newcommand{\gamonehs}{\hat{\gamma}_{1s}}
\newcommand{\bt}{{\bf t}}
\newcommand{\Igamone}{I_{\gam_1}}
\newcommand{\phigamone}{\varphi_{\gamonehs}}
\newcommand{\xigamone}{\xi_{\gamonehs}}
\newcommand{\ysgamone}{y_{s,\gamonehs}}
\newcommand{\gamtwohs}{\hat{\gamma}_{2s}}
\newcommand{\mgamhs}{|\hat{\gamma}_s|}
\newcommand{\dum}{s}
\newcommand{\xidum}{\xi_{\dum}}
\newcommand{\phidum}{\varphi_{\dum}}

\title[Statistics of Cosmic Shear]
{Analytical Predictions for Statistics of Cosmic Shear: 
Tests Against Simulations}

\author[P.Valageas et al.]
{Patrick Valageas$^{1}$,  
Andrew J. Barber$^{2}$, Dipak Munshi$^{3,4}$\\
$^{1}$ Service de Physique Th\'eorique, 
CEA Saclay, 91191 Gif-sur-Yvette, France \\
$^{2}$Astronomy Centre, University of Sussex, Falmer, Brighton, BN1 9QJ,
United Kingdom\\
$^{3}$Institute of Astronomy, Madingley Road,
Cambridge, CB3 OHA, United Kingdom\\
$^{4}$Astrophysics Group, Cavendish Laboratory, Madingley Road, 
Cambridge CB3 OHE, United Kingdom\\
}

\begin{document}
\maketitle

\begin{abstract}

Weak gravitational lensing surveys are rapidly becoming important
tools to probe directly the mass density fluctuations in the universe
and its background dynamics. Earlier studies have shown that it is
possible to model the statistics of the convergence field on small
angular scales by suitably modeling the statistics of the underlying
density field in the highly non-linear regime. We extend such methods
to model the complete probability distribution function, PDF, of the
shear as a function of smoothing angle. Our model relies on a simple 
hierarchical {\em Ansatz} for the
behaviour of the higher-order correlations in the density field. We
compare our predictions with the results of numerical
simulations and find excellent agreement for different cosmological
scenarios. Our method provides a new way to study the evolution of
non-Gaussianity in gravitational clustering and should help to break the 
degeneracies in parameter estimation based on analysis of the power 
spectrum alone.

\end{abstract}

 
%\maketitle 

\section{Introduction}

% recent detection

Recent detection of weak lensing (see, e.g., Bartelmann \& Schneider,
2001, for a review) due to the large-scale structure in the universe
has presented a new opportunity to study cosmology (see, e.g., Bacon,
Refregier \& Ellis, 2000, Hoekstra et al., 2002,
Van Waerbeke et al., 2000, and Van Waerbeke et al., 2002). With a
tight control on various systematics these studies were able to relate
the weak distortions in the images of high-redshift galaxies with the
intervening large-scale mass density inhomogeneities in the
universe. Such studies which are statistical in nature (as opposed to
strong lensing studies) provide us with a unique way to probe the
statistical properties of the underlying density
distribution. Traditionally, the study of gravitational clustering in
the quasi-linear and non-linear regimes was done by analyzing galaxy
catalogues and were plagued by the question of bias associated with
the galaxy distribution. Now, therefore, one needs a prescription for
galaxy bias to relate the galaxy distribution to the underlying dark
matter distribution.  Gravitational lensing allows us to bypass this
problem as it directly probes the mass distribution.  Pioneering works
in this direction were done by Blandford et al. (1991),
Miralda-Escud\'{e} (1991), and Kaiser (1992), based on the earlier
work by Gunn (1967).

A key tool for the understanding of cosmological weak lensing has been
provided by the development of numerical techniques to follow null
geodesics in a perturbed model universe. Such numerical studies
typically employ $N$-body simulations through which ray-tracing
experiments are conducted (see, e.g., Schneider \& Weiss, 1988,
Jaroszynski et al., 1990, Lee \& Paczynski, 1990, Jaroszynski,
1991, Babul \& Lee, 1991, and Blandford
et al., 1991). Building on the earlier work of Wambsganss et al. (1995
and 1997), detailed numerical studies were done by Wambsganss, Cen \&
Ostriker (1998). Other recent studies using ray-tracing have been
conducted by Premadi, Martel \& Matzner (1998), Van Waerbeke,
Bernardeau \& Mellier (1999) and White \&
Hu (2000). Recently, Couchman, Barber \& Thomas (1999) have developed
a new technique based on computing the full three-dimensional shear at
locations along the lines of sight. Their method combines the computed
shear matrices to produce the required lensing statistics for sources
at the selected redshift. The method is complimentary to ray-tracing
and has been fully implemented. Excellent agreement between the
results obtained using Couchman et al.'s (1999) method and analytical
predictions have been reported by Barber (2002) and Barber \& Taylor
(2002).

On the other hand, analytical approaches have also been developed to
elucidate the link between the statistical distortion of distant
sources by weak lensing and the properties of the large-scale
structure of the universe. Most of these analytical works have
focussed on large smoothing angles where perturbative calculations are
valid (e.g., Villumsen, 1996, Bernardeau et al., 1997,
Jain \& Seljak, 1997, Kaiser, 1998, Van Waerbeke, Bernardeau \&
Mellier, 1999, and Schneider et al., 1998).  Small angular scales are
more difficult to handle since they probe the non-linear regime where
the density field can no longer be described through perturbative
expansions which start behaving badly (e.g., Valageas, 2002b).  Thus,
for ongoing surveys with a small sky coverage perturbative results are
unlikely to be sufficient. Therefore, we need theoretical predictions
at small angular scales where the signal will dominate the noise.

% modeling non-linearity

Unfortunately, we lack a complete understanding of gravitational
clustering in the highly non-linear regime since it has not been
possible so far to solve the Vlasov-Poisson system which describes the
collisionless gravitational dynamics. Nevertheless, several
phenomenological models have been developed to describe the non-linear
density field. An interesting family of such models, often called
``hierarchical models'', assumes a tree hierarchy for many-body
correlation functions. Then, each model is specified by the way it
assigns weights to trees of the same order but of different topologies
(Fry, 1984, Schaeffer, 1984, Bernardeau \& Schaeffer, 1992, and
Szapudi \& Szalay, 1993 and 1997). Note that the evolution of the
two-point correlation function in all such approximations is left
arbitrary. On the other hand, independent studies by various authors
(Hamilton et al., 1991, and Peacock \& Dodds, 1994 and 1996) suggest
an accurate fitting formula for the evolution of the two-point
correlation function based on simple non-local scaling
arguments. Combining a hierarchical {\em Ansatz} with such a non-local
scaling for the two-point correlation one obtains a complete
statistical description of gravitational clustering in the highly
non-linear regime (e.g., Valageas \& Schaeffer, 1997, Munshi et al.,
1999a, Munshi, Coles \& Melott, 1999b,c, and Munshi, Melott \& Coles
1999d).

Such an approach was initiated by Hui (1999) for the study of the
three point correlation function for convergence maps and was later
extended to compute the full probability distribution function
(Valageas, 2000a,b, and Munshi \& Jain, 2000, 2001), the
associated bias (Munshi, 2000) and the cumulant correlators associated
with such distributions (Munshi \& Jain, 2000). See also Munshi \&
Wang (2003) for the extension of these studies to cosmological
scenarios with dark energy.  Next, such studies were extended to
handle the compensated filter appropriate for the aperture-mass,
$\Map$, which is more useful in weak lensing surveys (Bernardeau \&
Valageas, 2000). Error estimations were also performed based on such a
hierarchical {\em Ansatz} to optimize survey strategies (Munshi \&
Coles, 2003), while implications for Type Ia Supernov\ae~studies were
carried out by Valageas (2000a) and Wang, Holz \& Munshi (2002).

% recent studies.

So far most analytical calculations have focussed on convergence maps
where analytical results are simplified due to the fact that the
convergence is a scalar. Similar progress in the case of the shear was
lacking due to the spinorial nature of shear fields. However from an
observational point of view, shear maps are a direct outcome of suveys
whereas map-making for the convergence field can be problematic due to
the non-trivial topology of the survey area. Therefore, extending
previous studies of weak gravitational lensing based on hierarchical
models, we introduce in this article a model, which we also call a
``stellar model,'' to evaluate the probability distribution function,
PDF, of the shear. We test our analytical predictions against
numerical simulations and we show that our simple model provides a
good description of the shear PDF at small angular scales. Our
approach is complimentary to studies related to lower order statistics
of the shear field (Takada \& Jain, 2003, Zaldarriaga \&
Scoccimarro, 2003, Schneider \& Lombardi, 2003, and Bernardeau, Mellier 
\& van Waerbeke, 2002) which also aim to detect non-Gaussianity
from ongoing surveys.

This paper is organized as follows. Section $2$ briefly describes the
weak gravitational lensing effects as measured through the
convergence, $\kappa$, and the shear, $\gamma$. In Section $3$, we
recall the relationship between PDFs and cumulants (which are the
basic tools of our analytical method) and we present our model for the
many-body density correlations. This allows us to compute in Section
$4$ the PDFs for the shear components and for the shear
modulus. Section $5$ contains a brief explanation of the method used
to obtain the lensing statistics from the numerical simulations. In
Section $6$, we make the comparison between these numerical results
and our anaytical predictions, and we conclude in Section $7$ by
discussing our results and future work.

\section{Distortions induced by weak gravitational lensing}
\label{Distortions induced by weak gravitational lensing}

\subsection{Shear tensor}
\label{Shear tensor}

The gravitational lensing effects produced by density fluctuations along the 
trajectory of a photon lead to an apparent displacement of the source and to 
a distortion of the image. Thus, light coming from a direction ${\vec \theta}$ 
is deflected by a small angle $\delta {\vec \theta}$. However, the observable 
quantities are not the displacements $\delta {\vec \theta}$ themselves but 
the distortions induced by these deflections, which are given by the symmetric 
shear matrix (e.g., Jain, Seljak \& White, 2000):
\beq
\Phi_{i,j} = \frac{\pl \delta \theta_{i}}{\pl \theta_{j}} = -2 \int_0^{\chi_s}
 \d\chi \; \frac{\De(\chi) \De(\chi_s-\chi)}{\De(\chi_s)} \; \nabla_i 
\nabla_j \phi(\chi) .
\label{Phi}
\eeq
Here $\chi$ is the radial comoving coordinate (and $\chi_s$ corresponds to 
the redshift, $z_s$, of the source):
\beq
\d\chi = \frac{\frac{c}{H_0} \; \d z}{\sqrt{\Ol+(1-\Om-\Ol)(1+z)^2+
\Om(1+z)^3}} ,
\label{chi}
\eeq
while the angular distance $\De$ is defined by:
\beq
\De(\chi) = \frac{ \frac{c}{H_0} \sin_K \left( \mid 1-\Om-\Ol \mid^{1/2} 
H_0 \; \chi/c \right) } {\mid 1-\Om-\Ol \mid^{1/2}} ,
\label{De}
\eeq
where $\sin_K$ means the hyperbolic sine, sinh, if $(1-\Om-\Ol) > 0$,
or sine if $(1-\Om-\Ol) < 0$; if $(1-\Om-\Ol) = 0$, then $\De(\chi) = \chi$.
The gravitational potential, $\phi$, is related to the fluctuations of the 
density contrast, $\delta$, by Poisson's equation:
\beq
\Delta \phi = \frac{3}{2} \Om \; \frac{H_0^2}{c^2} \; (1+z) \; \delta 
\hspace{0.3cm} \mbox{with} \hspace{0.3cm} \delta(\bx) = 
\frac{\rho(\bx)-\rhob}{\rhob} ,
\label{Poisson}
\eeq
where $\rhob$ is the mean density of the universe. In eq.(\ref{Phi}) we used 
the weak lensing approximation; the derivatives $\nabla_i \nabla_j \phi(\chi)$
of the gravitational potential are computed along the unperturbed trajectory 
of the photon. This assumes that the components of the shear tensor are small 
but the density fluctuations $\delta$ can be large (Kaiser, 1992). The shear 
tensor $\Phi_{i,j}$ is usually decomposed into its trace, $\kappa$, and the 
shear components $\gam_1$, $\gam_2$, defined by:
\beq
\kappa = - \frac{\Phi_{1,1}+\Phi_{2,2}}{2} 
\label{defkappa}
\eeq
and
\beq
\gam_1 = - \frac{\Phi_{1,1}-\Phi_{2,2}}{2} \hspace{0.2cm} , \hspace{0.2cm} 
\gam_2 = - \Phi_{1,2} \hspace{0.2cm} , \hspace{0.2cm} 
\gam = \gam_1 + i \; \gam_2 .
\label{defgam}
\eeq
Note that since the shear components, $\gam_1$ and $\gam_2$, and their signs 
can be exchanged through appropriate rotations of the coordinate axes the 
joint probability distribution function (PDF) $\cP(\gam_1,\gam_2)$ obeys:
\beq
\cP(\gam_1,\gam_2) = \cP(|\gam_1|,|\gam_2|) = \cP(|\gam_2|,|\gam_1|) .
\label{sym1}
\eeq

\subsection{Convergence: $\kappa$}
\label{Convergence kappa}

Using eq.(\ref{Phi}) and eq.(\ref{Poisson}) one can show (Bernardeau et 
al., 1997, and Kaiser, 1998) that the convergence along a given line 
of sight is:
\beq
\kappa \simeq \frac{3\Om}{2} \int_0^{\chi_s} \d\chi \; w(\chi,\chi_s) 
\; \delta(\chi) ,
\label{kappa}
\eeq
with:
\beq
w(\chi,\chi_s) = \frac{H_0^2}{c^2} \; \frac{\De(\chi) \De(\chi_s-\chi)}
{\De(\chi_s)} \; (1+z) ,
\label{w}
\eeq
where $z$ corresponds to the radial distance $\chi$. Thus the convergence, 
$\kappa$, can be expressed very simply as a function of the 
density field; it is merely an average of the local density contrast along 
the line of sight. Then, we can see from eq.(\ref{kappa}), that there is a 
minimum value, $\kappamin(z_s)$, for the convergence of a source located at 
redshift $z_s$, which corresponds to an ``empty'' beam between the source 
and the observer ($\delta=-1$ everywhere along the line of sight):
\beq
\kappamin = - \frac{3\Om}{2} \int_0^{\chi_s} \d\chi \; w(\chi,\chi_s) .
\label{kappamin}
\eeq
Following Valageas (2000a, b) it is convenient to define the ``normalized'' 
convergence, $\kappah$, by:
\beq
\kappah = \frac{\kappa}{|\kappamin|} = \int_0^{\chi_s} \d\chi \; \wh \; 
\delta , \hspace{0.2cm} \mbox{with} \hspace{0.2cm} 
\wh=\frac{w(\chi,\chi_s)}{\int_0^{\chi_s} \d\chi \; w(\chi,\chi_s)} ,
\label{kappah}
\eeq
which obeys $\kappah \geq -1$. Here we introduced the ``normalized selection 
function,'' $\wh(\chi,\chi_s)$. One interest of working with normalized 
quantities like $\kappah$ is that most of the cosmological dependence 
(on $\Om,\Ol$ and $z_s$) and the projection effects are encapsulated within 
$\kappamin$, while the statistics of $\kappah$ (e.g., its PDF) mainly probe
the deviations from Gaussianity of the density field which arise from the
non-linear dynamics of gravitational clustering. If one smoothes the 
observations with a top-hat window in real space of small angular 
radius, $\theta_s$, one rather considers the filtered normalized convergence 
$\kaphs$ (where the subscript ``s'' refers to ``smoothed''):
\beq
\kaphs = \int_0^{\theta_s} \frac{\d {\vec \vartheta}}{\pi \theta_s^2} 
\int_0^{\chi_s} \d\chi \; \wh(\chi,\chi_s) \; 
\delta \left( \chi, \De {\vec \vartheta} \right) .
\label{kapthe}
\eeq
Here ${\vec \vartheta}$ is a vector in the plane perpendicular to the line 
of sight (we restrict ourselves to small angular windows) over which we 
integrate within the disk $|{\vec \vartheta}| \leq \theta_s$;
we note this by 
the short notation $\int_0^{\theta_s}$. Thus $\chi$ is the radial coordinate,
while $\De {\vec \vartheta}$ is the two-dimensional vector of transverse 
coordinates. Eq.(\ref{kapthe}) clearly shows that the convergence $\kaphs$ is 
actually an average of the density contrast over the cone of angular radius 
$\theta_s$. 

In the following, it will be convenient to work in Fourier space. Thus, we 
define the Fourier transform of the density contrast by:
\beq
\delta({\bx}) = \int \d\bk \; e^{i \bk.\bx} \; \delta(\bk)
\label{deltak}
\eeq
where $\bx$ and $\bk$ are comoving coordinates. Then, eq.(\ref{kapthe}) also 
reads:
\beq
\kaphs = \int_0^{\chi_s} \d\chi \; \wh(\chi,\chi_s) \int \d\bk \; 
e^{i \kpar \chi} \; W(\kperpDt) \; \delta( {\bf k} ) ,
\label{kappak}
\eeq
where $\kpar$ is the component of $\bk$ parallel to the line of sight
and $\kperp$ is the two-dimensional vector formed by the components of 
$\bk$ perpendicular to the line of sight. Here we introduced the Fourier 
form $W(\kperpDt)$ of the real-space top-hat filter of angular radius 
$\theta_s$:
\beq
W(\kperpDt) = \int_0^{\theta_s} \frac{\d {\vec \vartheta}}{\pi \theta_s^2} 
\; e^{i \kperp . \De {\vec \vartheta}} = \frac{2 J_1(\kperpDt)}{\kperpDt} ,
\label{Wk}
\eeq
where $J_1$ is the Bessel function of the first kind of order 1. If we 
choose another filter (e.g., a Gaussian window rather than a top-hat) the 
expression (\ref{kappak}) remains valid and we simply need to use the relevant 
Fourier window $W(\kperpDt)$.

\subsection{Shear: $\gam$}
\label{Shear}

For the shear, $\gam$, defined in eq.(\ref{defgam}), we can obtain expressions 
which are similar to eqs.(\ref{kapthe})-(\ref{kappak}). Thus, from 
eq.(\ref{Phi}) and eq.(\ref{defgam}) we obtain in Fourier space for the 
normalized shear $\gamh = \gam/|\kappamin|$:
\beq
\gamh = \int_0^{\chi_s} \d\chi \; \wh(\chi,\chi_s) \int \d\bk \; 
e^{i \bk.\bx} \frac{k_1^2 - k_2^2 + 2 i k_1 k_2}{k_1^2 + k_2^2} \delta(\bk)
\label{gamh1}
\eeq
where $k_1$ and $k_2$ are the components of $\bk$ along the two
orthogonal axes perpendicular to the line of sight which also define
the directions (1, 2) used in Subsection~\ref{Shear tensor} for the components 
$\Phi_{i,j}$ of the shear tensor. In eq.(\ref{gamh1}) we used again the 
small-angle approximation. Of course, for the smoothed shear $\gamhs$ we get, 
in a similar fashion to eq.(\ref{kappak}):
\beqa
\gamhs & = & \int_0^{\chi_s} \d\chi \; \wh \int \d\bk \; e^{i \kpar \chi} 
\; \frac{k_1^2 - k_2^2 + 2 i k_1 k_2}{k_1^2 + k_2^2} \nonumber \\ 
& & \times W(\kperpDt) \; \delta(\bk) .
\label{gamhk}
\eeqa
Going back to real-space, we also obtain from eq.(\ref{gamhk}) for a top-hat 
filter the expression:
\beq
\gamhs = - \int_0^{\chi_s} \d\chi \; \wh \int_{\theta_s}^{\infty} 
\frac{\d \vartheta}{\vartheta} \int_0^{2\pi} \frac{\d\alpha}{\pi} \; 
e^{i 2 \alpha} \; \delta(\chi, \De {\vec \vartheta} )
\label{gamreal}
\eeq
where ${\vec \vartheta}$ is a vector in the plane perpendicular to the line 
of sight of length $\vartheta$ and which makes the polar angle, $\alpha$, with 
the 1-axis (the 2-axis has $\alpha=\pi/2$). Note that we integrate over 
${\vec \vartheta}$ over {\it all space outside of the disc of radius 
$\theta_s$}. Thus, the comparison with eq.(\ref{kapthe}) shows that while the 
smoothed convergence only depends on the matter {\it within} the cone formed 
by the angular window, $\theta_s$, the smoothed shear only depends on the 
matter {\it outside} this cone.

\section{Cumulants, PDFs and models for the many-body density correlations}
\label{Cumulants and PDFs}

In this paper, we wish to evaluate the probability distribution
function, PDF, of the shear, $\gam$. To this order, following the
approach developed in Valageas (2000a, b) and Munshi \& Jain (2000,
2001), we shall first compute the moments (or more precisely the
cumulants) of the shear and then obtain the associated PDF. Therefore,
we first recall the standard relationship between the PDF and the
moment and cumulant generating functions.

\subsection{Generating functions}
\label{Generating functions}

The statistical properties of a random variable $\dum$ can be obtained from 
its moment or cumulant generating functions, which are widely used in 
statistics (see also Balian \& Schaeffer, 1989). Thus, the PDF $\cP(\dum)$ 
can be derived from the generating function, $\psi(y)$, defined from the 
moments, $\lag \dum^p\rag$, (provided they are finite) by: 
\beq
\psi(y) = \sum_{p=0}^{\infty} \frac{(-1)^p}{p!} \; \lag\dum^p\rag \; y^p
\label{psi1}
\eeq
since we have the inverse Laplace transform:
\beq
\cP(\dum) = \inta \frac{\d y}{2\pi i} \; e^{\dum y} \; \psi(y) .
\label{Pmu1}
\eeq
Indeed, the generating function, $\psi(y)$, is simply the Laplace transform 
of the PDF $\cP(\dum)$:
\beq
\psi(y) = \int \d\dum \; e^{-\dum y} \; \cP(\dum) .
\label{psi2}
\eeq
One can check that expanding the exponential in eq.(\ref{psi2}) reproduces 
eq.(\ref{psi1}). If the moments, $\lag\dum^p\rag$, diverge one can 
still define 
$\psi(y)$ from eq.(\ref{psi2}) but the expansion at $y=0$ of the generating 
function becomes singular. In practice, one usually introduces the generating 
function, $\Phi(y)$, defined from the cumulants, $\lag\dum^p\rag_c$, by:
\beq
\Phi(y) = \sum_{p=1}^{\infty} \frac{(-1)^p}{p!} \; \lag\dum^p\rag_c \; y^p 
= \ln \left[ \psi(y) \right] .
\label{Phi1}
\eeq
As is well-know, $\Phi(y)$ is simply the logarithm of $\psi(y)$. For our 
purposes, we shall actually use the normalized generating function, 
$\varphi(y)$, defined by:
\beq
\varphi(y) = \sum_{p=1}^{\infty} \frac{(-1)^{p-1}}{p!} \; 
\frac{\lag\dum^p\rag_c}{\xidum^{p-1}} \; y^p \hspace{0.3cm} \mbox{with} 
\hspace{0.3cm} \xidum = \lag\dum^2\rag_c .
\label{phi1}
\eeq

The comparison of eq.(\ref{phi1}) with eq.(\ref{Phi1}) yields
$\psi(y)$ $=$ $\exp[-\varphi(y\xidum)/\xidum]$, so that eq.(\ref{Pmu1}) 
now reads:
\beq
\cP(\dum) = \inta \frac{\d y}{2\pi i \xidum} \; 
e^{[\dum y-\varphi(y)]/\xidum} .
\label{Pmu2}
\eeq

As seen in Section~\ref{Distortions induced by weak gravitational
lensing}, the convergence, $\kappa$, or the shear, $\gam$, are given
by a linear integral along the line of sight over the density
field. In other words, they can be seen as the superposition of
independent layers which each contribute to $\kappa$ or
$\gam$. Indeed, within the small-angle approximation the correlations
of the density field will only appear within the transverse
directions. Then, the direct calculation of the PDF of such a sum over
the redshift of the lenses would yield an infinite number of
convolution products which makes it intractable. By contrast, the
cumulant generating functions simply add when different layers are
superposed (since Laplace transforms change convolutions into ordinary
products). This property makes the generating function $\varphi(y)$ a
convenient tool to deal with projection effects along the line of
sight. This remark is the basis of the method introduced in Valageas
(2000a, b). Moreover, the expansion (\ref{phi1}) provides a simple way
to compute $\varphi(y)$ from the moments (or cumulants) of the random
variable one is interested in. This approach can be used for any
quantity which is linear over the density field, like the convergence,
$\kappa$, studied in Valageas (2000a, b), the aperture-mass, $\Map$,
investigated in details in Bernadeau \& Valageas (2000) or the shear,
$\gam$, we focus on in this paper.

\subsection{PDF for the density field}
\label{PDF for the density field}

It is clear that in order to derive the PDF of the shear, $\gam$, we need the 
properties of the underlying density field. A first measure of the 
statistical properties of the density field is given by the PDF 
$\cP(\delta_R)$ of the density contrast within spherical cells of radius $R$ 
and volume $V$:
\beq
\delta_R = \int_V \frac{\d\bx}{V} \; \delta(\bx) .
\label{deltaR1}
\eeq
Note that the cumulants, $\lag\delta_R^p\rag_c$, can also be written in terms 
of the many-body connected correlation functions, $\xi_p(\bx_1,..,\bx_p)$, 
as (e.g., Peebles, 1980):
\beq
p \geq 2 \; : \;\; \lag\delta_R^p\rag_c = \xib_p = \int_V 
\frac{\d\bx_1 .. \d\bx_p}{V^p} \; \xi_p (\bx_1,..,\bx_p) ,
\label{xibp}
\eeq
with:
\beq
\xi_p (\bx_1,..,\bx_p) = \lag \delta(\bx_1) .. \delta(\bx_p) \rag_c .
\label{xip}
\eeq
Next, we define the associated normalized generating function, $\varphi(y)$, 
introduced in eq.(\ref{phi1}) and its coefficients, $S_p$,  by:
\beq
\varphi(y) = \sum_{p=2}^{\infty} \frac{(-1)^{p-1}}{p!} \; S_p \; y^p , 
\hspace{0.3cm} S_p = \frac{\xib_p}{\xib_2^{\; p-1}} , 
\hspace{0.2cm} S_2 = 1 ,
\label{phideltaR}
\eeq
which also yields $\cP(\delta_R)$ through eq.(\ref{Pmu2}):
\beq
P(\delta_R) = \inta \frac{\d y}{2\pi i \xib_2} \; 
e^{[\delta_R y - \varphi(y)] /\xib_2} .
\label{PdeltaR}
\eeq
Since the convergence, $\kaphs$, is the average of the density
contrast, $\delta$, with a smooth positive weight over the relevant
cone, its cumulants can be estimated from those of the density field
in a robust and simple way, by neglecting their geometrical
dependence. Therefore, one merely uses the properties of the density
field, $\delta_R$, smoothed over spherical cells and does not take into
account the detailed angular dependence of the many-body correlation
functions. As shown in Valageas (2000b), this procedure is indeed sufficient
to recover the properties of the convergence. In fact, up to a good
accuracy, one may even express the PDF, $\cP(\kaphs)$, directly in
terms of the PDF, $\cP(\delta_R)$, of the density contrast, by using the
approximation $\phikap(y) \simeq \varphi(y)$ (see Valageas 2000b).
Here $\varphi$ (also defined in eq.(\ref{phideltaR})) is the normalized 
generating function associated with the density contrast $\delta_R$ at
the scale and redshift which contribute most to the signal, whereas $\phikap$ 
is the normalized generating function associated with the normalized
convergence, $\kaphs$, as defined by eq.(\ref{phi1}). This 
approximation clearly shows that the ``normalized convergence,'' $\kaphs$,
mainly probes the non-Gaussianities of the density field (i.e., its very 
structure) brought by the non-linear gravitational dynamics, while the overall
amplitude of the fluctuations (which gives the dependence on the cosmological
parameters and the source redshift) is encoded within $\kappamin$, as we
announced earlier, below eq.(\ref{kappah}).

\subsection{Models for the many-body density correlations}
\label{Models for the many-body density correlations}

For more intricate filters, this simple approximation is no longer
sufficient.  This is clearly the case for the shear, $\gam$, since
eq.(\ref{gamreal}) shows that the relevant volume is now the exterior
of the cone of radius $\theta_s$, and has a geometry which is
significantly different from spherical cells. More importantly, the
factor $e^{i2\alpha}$ yields a weight of mean zero and which is both
positive and negative and varies with the direction in the transverse
plane. This implies that one cannot evaluate the cumulants of $\gam$
from these moments of the density field over spherical cells; the
natural estimate would be zero, since the angular mean of the weight,
$e^{i2\alpha}$, vanishes. The same problem appears for the study of
the aperture mass, $\Map$, which also involves a compensated
filter. Therefore, we need an explicit model for the angular
dependence of the many-body correlations.

As shown in Bernardeau \& Valageas (2000), in the case of the aperture mass,
$\Map$, good results can be obtained at small angles by using a ``minimal 
tree-model'' for the connected correlations of the density field. Let us 
first recall that a general ``tree-model'' is defined by the property 
(Schaeffer, 1984, and Groth \& Peebles, 1977):
\beq
\xi_p(\bx_1, .. ,\bx_p) = \sum_{(\alpha)} Q_p^{(\alpha)} \sum_{t_{\alpha}} 
\prod_{p-1} \xi_2(\bx_i,\bx_j) 
\label{tree}
\eeq
where $(\alpha)$ is a particular tree-topology connecting the $p$ points 
without making any loop, $Q_p^{(\alpha)}$ is a parameter associated with the 
order of the correlations and the topology involved, $t_{\alpha}$ is a 
particular labeling of the topology, $(\alpha)$, and the product is made over 
the $(p-1)$ links between the $p$ points with two-body correlation functions. 
We show in Fig.1 the three topologies which appear within this framework for 
the 5-point connected correlation function.

\begin{figure}
\protect\centerline{\epsfysize = 3 cm \epsfbox{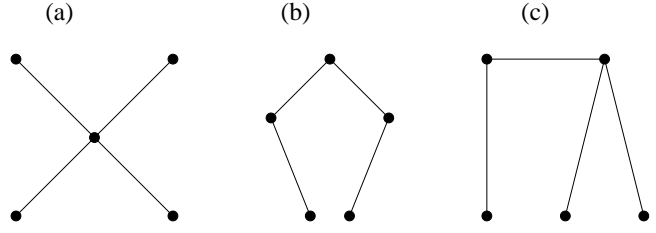}}
\caption{The various topologies one can build for the 5-point connected 
correlation function within the framework of a tree-model as in (\ref{tree}). 
Graph (a) is a ``stellar diagram'' (i.e., $(p-1)$ points are linked to a 
central point), while graph (b) is a ``snake diagram'' (i.e., one successively 
visits all points along one curve with only two end-points) and graph (c) 
is a ``hybrid diagram''.}
\end{figure}

\subsubsection{Minimal tree model}
\label{Minimal tree-model}

A peculiar case of the tree-models described by (\ref{tree}) is the 
minimal tree-model where the weights $Q_p^{(\alpha)}$ are given by 
(Bernardeau \& Schaeffer, 1992): 
\beq
Q_p^{(\alpha)} = \prod_{\mbox{vertices of } (\alpha)} \nu_q
\label{mintree}
\eeq
where $\nu_q$ is a constant weight associated to a vertex of the tree 
topology with $q$ outgoing lines. Then, for an arbitrary real-space 
filter, $F(\bx)$, which defines the random variable $\dum$ as:
\beq
\dum = \int \d\bx \; F(\bx) \; \delta(\bx) \hspace{0.4cm} \mbox{and} 
\hspace{0.4cm} \xidum= \lag \dum^2 \rag ,
\label{mu}
\eeq
it is possible to obtain a simple implicit expression for the normalized 
generating function, $\phidum(y)$ (see Bernardeau \& Schaeffer, 1992,
and Jannink \& Des Cloiseaux, 1987): 
\beqa
{\displaystyle \phidum(y)} & = & {\displaystyle y \int \d\bx \; F(\bx) \; 
\left[ \zeta[ \tau(\bx)] - \frac{\tau(\bx) \zeta'[\tau(\bx)]}{2} \right] } 
\label{implicitphi} \\ 
{\displaystyle \tau(\bx) } & = & {\displaystyle \frac{-y}{\xidum} 
\int \d\bx' \; F(\bx') \; \xi_2(\bx,\bx') \; \zeta'[\tau(\bx')] } 
\label{implicittau}
\eeqa
where the function $\zeta(\tau)$ is defined as the generating function for 
the coefficients $\nu_p$:
\beq
\zeta (\tau) = \sum_{p=1}^{\infty} \frac{(-1)^p}{p!} \; \nu_p \; \tau^p 
\hspace{0.4cm} \mbox{with} \hspace{0.4cm} \nu_1 = 1.  
\label{zeta1}
\eeq
Of course, the generating function, $\phidum(y)$, depends on the filter, $F$, 
which defines the variable, $\dum$. For the smoothed density contrast, 
$\delta_R$, defined in eq.(\ref{deltaR1}), the real-space filter, $F(\bx)$, is 
the top-hat of radius $R$ normalized to unity. In this case, a simple 
``mean field'' approximation which provides very good results 
(Bernardeau \& Schaeffer, 1992) is to integrate $\tau(\bx)$ over the volume 
$V$ in eq.(\ref{implicittau}) and then to approximate $\tau(\bx)$ by a 
constant $\tau$. This leads to the simple system: 
\beqa
\varphi(y) & = & y \left[ \zeta(\tau)-\frac{\tau\;\zeta'(\tau)}{2} \right] 
\label{phiMF} \\
\tau & = & - y \; \zeta'(\tau)
\label{tauMF}
\eeqa
Here and in the following we simply note the generating function associated 
with the density contrast $\delta_R$ as $\varphi(y)$ and we retain suffixes 
for the generating functions associated with other quantities 
(e.g., $\phigamone$ for the shear component, $\gamonehs$).

In the case of the aperture mass, $\Map$, one can still perform the
resummation (\ref{implicitphi})-(\ref{implicittau}), if the underlying
density field is described by such a minimal tree-model, but one can
no longer use the ``mean field'' approximation
(\ref{phiMF})-(\ref{tauMF}) because the 2-d compensated filter
$F({\vec \vartheta})$ now strongly depends on the polar distance
$|{\vec \vartheta}|$ (in the 2-d transverse plane). Such a study is
described in detail in Bernardeau \& Valageas (2000) where it is shown
that this minimal tree-model provides good results when compared with
numerical simulations.

Of course, we can apply the same procedure, developed in Bernardeau 
\& Valageas (2000) for $\Map$, to the shear, $\gam$. We simply need to use the 
corresponding real-space filter, $F$. From eq.(\ref{gamreal}) we see that 
this is $F({\vec \vartheta})= - H(\vartheta > \theta_s) \cos(2\alpha) 
/\pi\vartheta^2$ for $\gam_1$ and $F({\vec \vartheta})= 
- H(\vartheta > \theta_s) \sin(2\alpha) /\pi\vartheta^2$ for $\gam_2$, 
where $H$ is the Heaviside function with obvious notations. Then, the 
resummation (\ref{implicitphi})-(\ref{implicittau}) yields a two-dimensional 
non-linear implicit equation which is not very convenient for numerical 
purposes. Therefore, it is interesting in this case to investigate whether 
a simpler model for the many-body correlations of the density field could 
still provide good predictions for the PDF, $\cP(\gamma_1)$, of the shear 
component, $\gamma_1$.

\subsubsection{Stellar model}
\label{Stellar-model}

Thus, we introduce in this paper a ``stellar-model'', which is a particular 
case of the tree-models defined in (\ref{tree}), where we only keep the 
stellar diagrams (e.g., the graph (a) in Fig.1 for the 5-point connected 
correlation). Thus, the $p-$point connected correlation $\xi_p$ of the density 
field can now be written as:
\beq
\xi_p(\bx_1, .. ,\bx_p) = \frac{\tS_p}{p} \; \sum_{i=1}^p \prod_{j\neq i} 
\xi_2(\bx_i,\bx_j) .
\label{xistar1}
\eeq
The advantage of the stellar-model (\ref{xistar1}) is that it leads to 
very simple calculations in Fourier space. Thus, let us define the 
power-spectrum, $P(k)$, of the density contrast by:
\beq
\lag \delta(\bk_1) \delta(\bk_2) \rag = \delta_D(\bk_1+\bk_2) \; P(k_1) ,
\label{Pk}
\eeq
where $\delta_D$ is Dirac's distribution, whence:
\beq
\xi_2(x) = \int \d\bk \; e^{i\bk.\bx} \; P(k) .
\label{xiPk}
\eeq
Then, eq.(\ref{xistar1}) reads in Fourier space:
\beq
\lag \delta(\bk_1) .. \delta(\bk_p) \rag_c = \frac{\tS_p}{p} \; 
\delta_D(\bk_1+..+\bk_p) \sum_{i=1}^p \prod_{j\neq i} P(k_j) .
\label{corrstar1}
\eeq
Of course, the Dirac factor, $\delta_D(\bk_1+..+\bk_p)$, simply translates the 
fact that the many-body correlations are invariant through translations. 
Therefore, in the following we shall use eq.(\ref{corrstar1}) for the 
connected correlations of the density field which appear in the calculation 
of the cumulants of the shear, $\gam$.

\subsection{Amplitude of the density correlations}
\label{Amplitude of the density correlations}

The stellar-model introduced in Section~\ref{Stellar-model} provides
the angular dependence of the many-body correlations of the density
field.  However, we still need to determine their overall amplitude,
that is the non-linear power-spectrum, $P(k)$, and the coefficients,
$\tS_p$, which enter eqs.(\ref{xistar1})-(\ref{corrstar1}). First, we
obtain the non-linear evolution of the power-spectrum from the
prescription given by Peacock \& Dodds (1996), which is also a good
fit to numerical simulations.  Second, we determine the coefficients,
$\tS_p$, as follows. We characterize the amplitude of the density
correlations at scale $R$ and redshift $z$ through the parameters
$S_p$ defined in eq.(\ref{phideltaR}), which give the amplitude of the
many-body correlations over spherical cells of radius $R$.  As seen
from eq.(\ref{phideltaR})-(\ref{PdeltaR}) they also define the PDF,
$\cP(\delta_R)$. We could obtain these coefficients, $S_p$, through a
parameterization of their generating function, $\varphi(y)$. However,
we prefer to specify $\varphi(y)$ through a function $\zeta(\tau)$
which is defined by the implicit system
(\ref{phiMF})-(\ref{tauMF}). Note that this procedure is merely a
convenient parameterization of $\varphi(y)$ and it does not assume by
itself a minimal tree-model. The results recalled in
Section~\ref{Minimal tree-model} simply mean that in case the actual
many-body correlations exactly obey a minimal tree-model, the function
$\zeta(\tau)$ defined from eqs.(\ref{phiMF})-(\ref{tauMF}) could also
be interpreted as a good approximation to the (slightly different)
function $\zeta_{\nu}(\tau)$ defined in eq.(\ref{zeta1}) as the
generating function of the vertices $\nu_p$.

The reason we prefer to parameterize the coefficients, $S_p$, through 
$\zeta(\tau)$ rather than directly through $\varphi(y)$ is that this 
procedure was already checked to provide good results in previous works. 
Thus, it was seen in Bernardeau \& Schaeffer (1992) to give good predictions 
for the galaxy and matter correlations, while in Bernardeau \& Valageas (2000) 
it was shown to yield good predictions for the PDF, $\cP(\Map)$, of the 
aperture-mass. Since the aperture-mass is closely related to the shear, 
$\gam$, we can expect this procedure to give good results for the PDF, 
$\cP(\gam)$, too. Note however that the calculations developed in 
Bernardeau \& Valageas (2000) used the minimal tree-model recalled in 
Section~\ref{Minimal tree-model} rather than the stellar-model of 
Section~\ref{Stellar-model} for the angular dependence of the many-body 
correlations. Nevertheless, we checked numerically that both models actually 
give close results for $\cP(\Map)$. Thus, as in 
Bernardeau \& Valageas (2000) we parameterize $\zeta(\tau)$ as:
\beq
\zeta(\tau)= \left (1+\frac{\tau}{\kappa}\right)^{-\kappa} - 1 ,
\label{zetadef}
\eeq
where we have kept the usual notation, $\kappa$ (not to be confused with the 
convergence), for the free parameter which enters the definition of 
$\zeta(\tau)$. 

Next, we choose the value of this parameter, $\kappa$, so as to reproduce the 
skewness, $S_3$, in both the highly non-linear and quasi-linear regimes, 
using the 
relation:
\beq
\kappa= \frac{3}{S_3-3} .
\label{S3kappa}
\eeq
For $S_3>3$ we have $\kappa>0$ while for $0<S_3<3$ we have $\kappa<-1$.
In the limit $S_3 \rightarrow 3$ the function $\zeta(\tau)$ defined in
eq.(\ref{zetadef}) goes to $\zeta(\tau) \rightarrow e^{-\tau}-1$ (and
$|\kappa| \rightarrow \infty$).
We take for the skewness in the highly non-linear regime, $S_3^{\rm NL}$, 
the prediction of HEPT (Scoccimarro \& Frieman, 1999) and for the quasi-linear 
regime, $S_3^{\rm QL}$, the exact result obtained from perturbative theory:
\beq
S_3^{\rm NL} = 3 \frac{4-2^n}{1+2^{n+1}} , \hspace{0.3cm} 
S_3^{\rm QL} = \frac{34}{7} - (n+3) .
\label{S31}
\eeq
Here $n$ is the local slope of the linear power-spectrum at the typical 
wavenumber $k_s$ probed by the observations at redshift $z$.
From eq.(\ref{kappak}) and eq.(\ref{gamhk}), we define this typical 
wavenumber as:
\beq
k_s(z) = \frac{1}{\De(z)\theta_s} .
\label{ks}
\eeq
Finally,
we introduce the power, $\Delta^2(k)$, per logarithmic interval defined as:
\beq
\Delta^2(k,z) = 4 \pi k^3 P(k,z) .
\label{Delta2k}
\eeq
Then, for intermediate regimes defined by $1 < \Delta^2(k_s,z) < 
\Delta_{\rm vir}(z)$, we use the simple linear interpolation:
\beq
S_3(z) = S_3^{\rm QL} + \frac{\Delta^2(k_s,z) -1}{\Delta_{\rm vir}(z)-1} 
\left( S_3^{\rm NL} - S_3^{\rm QL} \right) .
\label{S32}
\eeq
For $\Delta^2(k_s,z)<1$ (i.e. the quasi-linear regime), we take 
$S_3=S_3^{\rm QL}$, while for $\Delta^2(k_s,z) > \Delta_{\rm vir}(z)$ 
(i.e., the highly non-linear regime), we take $S_3=S_3^{\rm NL}$. Here 
$\Delta_{\rm vir}$ is the density contrast at virialization given by the 
usual spherical collapse (thus $\Delta_{\rm vir} \sim 178$ for a 
critical-density universe). Indeed, the threshold $\Delta^2 > 
\Delta_{\rm vir}$ describes the highly non-linear regime where most of the 
matter at scale $1/k$ has collapsed into non-linear structures.

Finally, as can be seen from eq.(\ref{corrstar1}), the coefficients $S_p$ 
and $\tS_p$ are related by:
\beq
S_p = \tS_p \int_0^1 dt \; 3t^2 \; \left[ \frac{\int \frac{dk}{k} \; 
\Delta^2(k) F(kR) \frac{\sin(kRt)}{kRt}}{\int \frac{dk}{k} \; \Delta^2(k) 
F^2(kR)} \right]^{p-1}
\label{SptSp1}
\eeq
where we introduced the Fourier transform, $F(kR)$, of a 3-d top-hat of radius 
$R$:
\beq
F(kR) = \int \frac{\d\bx}{V} \; e^{i\bk.\bx} = 3 \; 
\frac{\sin(kR) - (kR) \cos(kR)}{(kR)^3} .
\eeq
However, in the following we shall use the simple approximation:
\beq
\tS_p \simeq S_p .
\label{SptSp2}
\eeq
Alternatively, we may define the function $\varphi(y)$ obtained from 
(\ref{phiMF})-(\ref{tauMF}) and our choice of $\zeta(\tau)$ as the 
generating function of the coefficients $\tS_p$, rather than $S_p$, through 
its Taylor expansion at $y=0$.

It is interesting to note that the implicit system (\ref{phiMF})-(\ref{tauMF}) 
appears naturally in the quasi-linear regime. Indeed, as shown in 
Bernardeau (1994) and Valageas (2002a), in this regime the generating 
function $\varphi(y)$ exactly obeys the relation (\ref{phiMF})-(\ref{tauMF}), 
where $\zeta(\tau)$ is closely related to the spherical dynamics for 
the non-linear 
density contrast. However, the latter is of the form (\ref{zetadef}) only 
in the limit $\Om \rightarrow 0$ with no smoothing (then, the parameter 
$\kappa$ is equal to $3/2$). Nevertheless, the form (\ref{zetadef}) still 
provides a good approximation for more general cases where $\kappa$ is 
adjusted so as to recover the exact skewness. Here we note that the system 
(\ref{phiMF})-(\ref{tauMF}) usually yields a branch cut along the negative 
real axis for $y<y_s$ with $y_s \sim -0.1$ for $\varphi(y)$; see 
Bernardeau \& Schaeffer (1992). For the function $\zeta(\tau)$ given in 
eq.(\ref{zetadef}) one gets (for $\kappa>0$ or $\kappa<-2$, that is 
$S_3>3/2$):
\beq
y_s = - \frac{\kappa}{\kappa+2} \left( \frac{\kappa+1}{\kappa+2} 
\right)^{\kappa+1}
\hspace{0.4cm} \mbox{and} \hspace{0.4cm} \tau_s= - \frac{\kappa}{\kappa+2}.
\label{ys}
\eeq
Such a singularity leads to an exponential cutoff for the PDF, 
$\cP(\delta_R) \sim e^{y_s \delta_R/\xib_2}$ at large densities. As shown 
in Valageas (2002a), in the case of the quasi-linear regime this is actually 
an artefact and one must go through $y_s$ up to a second branch for 
$\varphi(y)$, which yields a large density cut-off for $\cP(\delta_R)$ which 
is shallower than a simple exponential. However, it was seen by numerical 
computations in Valageas (2002a) that this feature does not change  
the shape of the PDF by much in the range of interest. Moreover, in this 
paper we 
are mostly interested in the non-linear regime (which largely dominates for 
$\theta \la 4'$ and $z_s =1$) where the singularity $y_s$ is meaningful and 
the PDF shows indeed an exponential cutoff (since it is actually defined from 
the coefficients $S_p$).

Finally, we must point out that the simple model developed above has no new 
free parameter. Indeed, the power-spectrum evolution is obtained from 
Peacock \& Dodds (1996), while all many-body correlations are defined by the 
stellar-model (\ref{xistar1}) with their amplitude given by the sole 
coefficient $S_3$ (i.e. the skewness of $\cP(\delta_R)$) which is exact in 
the quasi-linear regime and obtained from HEPT (Scoccimarro \& Frieman, 1999) 
in the non-linear regime. In the form given here, our model is obviously 
consistent with the ``stable-clustering {\em Ansatz}'' (e.g., Peebles 1980). 
However, it could incorporate possible deviations from the 
stable-clustering {\em Ansatz} through the non-linear power-spectrum, $P(k)$, 
or through an additional dependence on redshift and scale for the skewness, 
$S_3$, in the non-linear regime.

\begin{figure}
\protect\centerline{
\epsfysize = 3.5truein
%\epsfbox[27 75 477 564]
\epsfbox[25 147 588 715]
{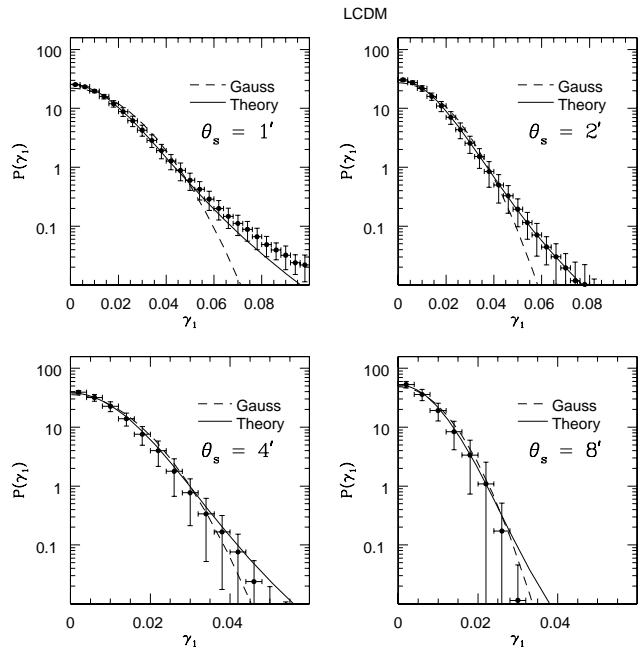} }
\caption{Probability distribution function, $\cP(\gamma_1)$, for the
LCDM cosmology, plotted as a function of $\gamma_1$. The smoothing
angle, $\theta_s$, varies from $1'$ (upper left panel) to $8'$ (lower
right panel). The data-points are the results from the numerical
simulations and the solid lines correspond to the analytical
predictions (\ref{phigam1}) and (\ref{Pgam1h1}) based on our 
stellar-model. The dashed lines show the Gaussian PDF with the same
variance. The source redshift is fixed at unity in each case.}
\end{figure}

\begin{figure}
\protect\centerline{
\epsfysize = 3.5truein
%\epsfbox[27 75 477 564]
\epsfbox[25 147 588 715]
{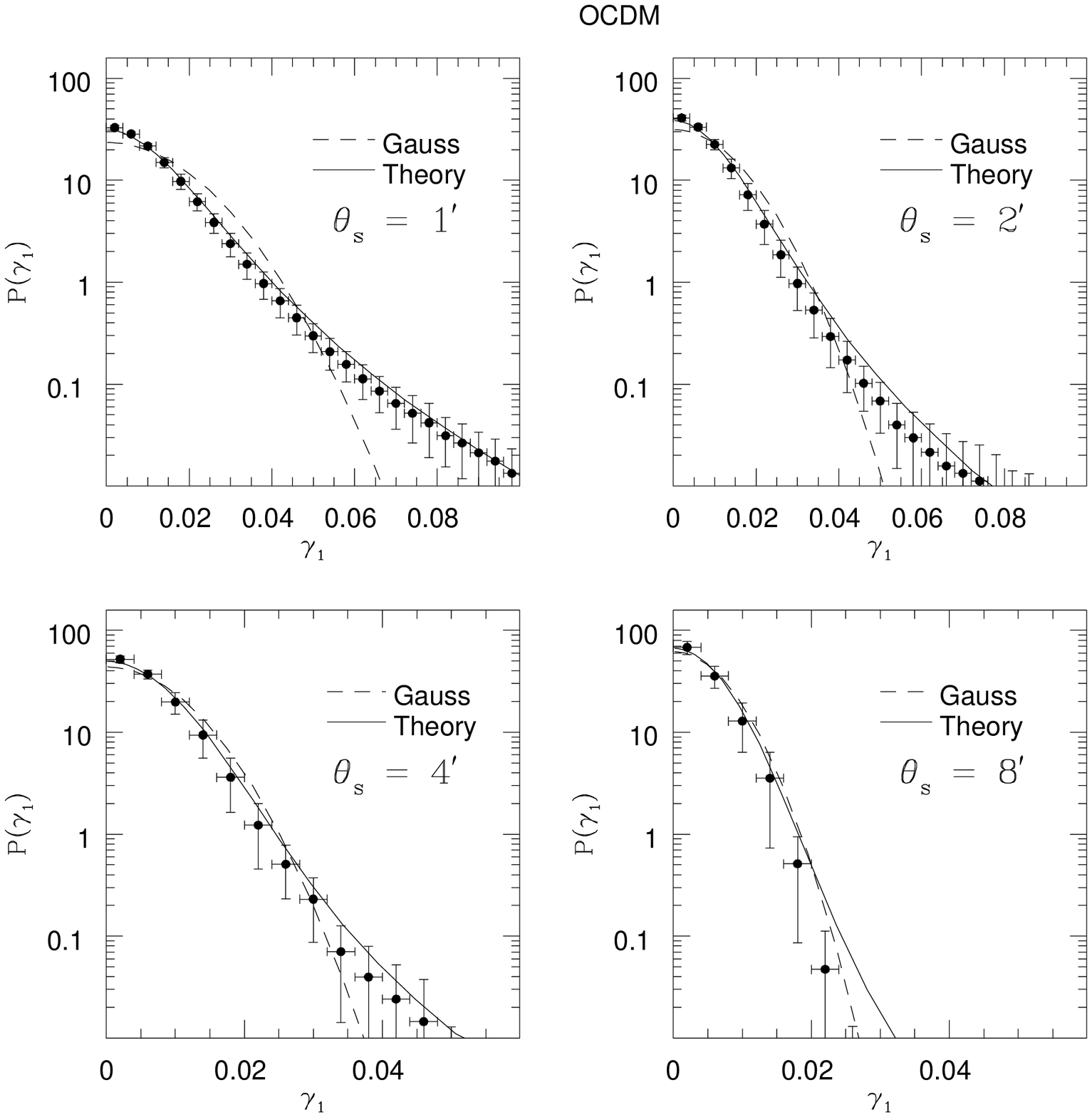} }
\caption{As for the previous Figure, but for the OCDM cosmology.}
\end{figure}

\section{Computation of the PDFs}

\subsection{PDF of the shear components, $\gam_1$ and $\gam_2$}
\label{PDF of the shear components}

We can now use the model developed in Section~\ref{Cumulants and PDFs} to 
describe the density field in order to compute the PDF, $\cP(\gam_1)$, of the 
shear component, $\gam_1$. First, we obtain from eq.(\ref{gamhk}):
\beq
\gamonehs = \int_0^{\chi_s} \d\chi \; \wh \int \d\bk \; e^{i \kpar \chi} \; 
\cos2\phi \; W(\kperpDt) \; \delta(\bk) ,
\label{gam1h1}
\eeq
where $\phi$ is the polar angle of the transverse vector $\kperp$. For the 
component $\gam_2$ we would have $\sin2\phi$ in place of $\cos2\phi$. Then, 
the cumulant of order $p$ reads:
\beqa
\lag \gamonehs^p \rag_c & = & \int_0^{\chi_s} \prod_{i=1}^{p} \d\chi_i \; 
\wh_i \int \prod_{j=1}^{p} \d\bk_j \; \cos2\phi_j \; 
W(k_{\perp j} \De \theta_s) \nonumber \\ 
& & \times \left( \prod_{l=1}^{p} e^{i k_{\parallel l} \chi_l} \right) 
\; \lag \delta(\bk_1) .. \delta(\bk_p) \rag_c .
\label{gam1h2}
\eeqa
Since the correlation length (beyond which the density correlations are 
negligible) is much smaller than the Hubble scale, $c/H(z)$ (where $H(z)$ 
is the Hubble constant at redshift $z$), we can simplify eq.(\ref{gam1h2}) as:
\beqa
\lefteqn{\!\!\! \lag \gamonehs^p \rag_c = \int_0^{\chi_s} \d\chi_1 \wh_1^p 
\int_{-\infty}^{\infty} \prod_{i=2}^{p} \d\chi_i \int \prod_{j=1}^{p} 
\d\bk_j \cos2\phi_j W(k_{\perp j} \De \theta_s) } \nonumber \\ 
& & \times \left( \prod_{l=1}^{p} e^{i k_{\parallel l} \chi_l} \right) \; 
\tS_p \; \delta_D(\bk_1+..+\bk_p) P(k_2) .. P(k_p) ,
\label{gam1h3}
\eeqa
where we used eq.(\ref{corrstar1}). Here the coefficients $\tS_p(z_1)$ are
evaluated at the typical wavenumber, $k_s(z_1)$, defined in eq.(\ref{ks}),
probed by the smoothed shear at the redshift $z_1$ along the line of sight.
Using the usual small-angle approximation
(i.e. $P(k_j) \simeq P(k_{\perp j})$), we can perform the integration over 
$\chi_2,..,\chi_p$ and $k_{\parallel 1},..,k_{\parallel p}$, which yields:
\beqa
\lag \gamonehs^p \rag_c & = & \int \frac{\d\chi}{2\pi} (2\pi\wh)^p 
\int \prod_{j=1}^{p} \d\bk_{\perp j} \; \cos2\phi_j \; 
W(k_{\perp j} \De \theta_s) \nonumber \\ 
& & \times \; \tS_p \; \delta_D(\bk_{\perp 1}+..+\bk_{\perp p}) 
P(k_{\perp 2}) .. P(k_{\perp p}) .
\label{gam1h4}
\eeqa
Next, we write the Dirac factor as:
\beq
\delta_D(\bk_{\perp 1}+..+\bk_{\perp p}) = \int \frac{\d\bt}{(2\pi)^2} 
\; e^{i\bt.(\bk_{\perp 1}+..+\bk_{\perp p})} ,
\label{Dirac1}
\eeq
so that eq.(\ref{gam1h4}) reads:
\beqa
\lefteqn{ \lag \gamonehs^p \rag_c = \int \frac{\d\chi}{2\pi} \; \tS_p \; 
(2\pi\wh)^p \int \frac{\d\bt}{(2\pi)^2} \int \d\bk_1 .. \d\bk_p } 
\nonumber \\ 
& & \!\!\!\!\!\! \times \prod_{j=1}^{p} \cos2\phi_j \; W(k_j \De \theta_s) 
e^{i t k_j \cos(\phi_j-\phi)} \times P(k_2) .. P(k_p)
\label{gam1h5}
\eeqa
where we have dropped the subscript, $\perp$, for the 2-d transverse 
wavenumbers 
$\bk_{\perp}$ and we have noted $\phi$ the polar angle of the vector $\bt$. 
The integration over the angles $\phi_j$ yields:
\beq
\int_0^{2\pi} \d\phi_j \; \cos2\phi_j \; e^{i t k_j \cos(\phi_j-\phi)} 
= - 2\pi \cos2\phi \; J_2(t k_j)
\eeq
so that eq.(\ref{gam1h5}) becomes:
\beqa
\lefteqn{ \lag \gamonehs^p \rag_c = \int_0^{\chi_s} \d\chi \; \tS_p \; \wh^p 
\int_0^{\infty} \d t \; t (\De \theta_s)^2 \int_0^{2\pi} \frac{\d\phi}{2\pi} 
\; (-\cos2\phi)^p } \nonumber \\ 
& & \times \int_0^{\infty} \d k_1 \; k_1 W(k_1 \De \theta_s) 
J_2(t k_1 \De \theta_s) \; \Igamone(t,\chi)^{p-1}
\label{gam1h6}
\eeqa
where we have made the change of variable $t \rightarrow t\De\theta$ and we 
have introduced $\Igamone(t,\chi)$ defined by:
\beqa
\Igamone(t,\chi) & = & 4 \pi^2 \int_0^{\infty} \d k \; k W(k \De \theta_s) 
J_2(t k \De \theta_s) P(k) \nonumber \\ 
& = & \pi \int_0^{\infty} \frac{\d k}{k} \; \frac{\Delta^2(k,z)}{k} \; 
W(k \De \theta_s) J_2(t k \De \theta_s) .
\label{Ig1}
\eeqa
For the angular top-hat filter, $W$, given by eq.(\ref{Wk}) in Fourier space, 
we can perform the integration over $k_1$ in eq.(\ref{gam1h6}), using the 
standard properties of Bessel functions, and we obtain:
\beq
\lag \gamonehs^p \rag_c = \int_0^{\chi_s} \d\chi \int_1^{\infty} 
\frac{\d t}{t} \int_0^{2\pi} \frac{\d\phi}{\pi} \; \tS_p \; (\wh\cos2\phi)^p 
\Igamone^{p-1} .
\label{gam1h7}
\eeq
Here we have dropped the minus sign in front of the factor $\cos2\phi$ because 
all odd moments vanish, as seen by integration over $\phi$. Thus, we recover 
the fact that the PDF, $\cP(\gam_1)$, is even, which was obvious from the 
definition of the shear since there is no preferred direction in the system. 

Finally, the expression (\ref{gam1h7}) allows us to obtain the generating 
function $\phigamone(y)$ of the shear component $\gamonehs$, defined as in 
eq.(\ref{phi1}), using the resummation (\ref{phideltaR}) (where we replace 
$S_p$ by $\tS_p$; see eq.(\ref{SptSp2})):
\beqa
\phigamone(y) & = & \int_0^{\chi_s} \d\chi \int_1^{\infty} \frac{\d t}{t} 
\int_0^{2\pi} \frac{\d\phi}{\pi} \; \frac{\xigamone}{\Igamone} \nonumber \\ 
& & \times \; \varphi \left( y \cos2\phi \frac{\Igamone}{\xigamone} 
\wh ; z \right) .
\label{phigam1}
\eeqa
Here we have introduced the variance $\xigamone=\lag \gamonehs^2 \rag$. From 
eq.(\ref{gam1h7}) it is given by:
\beqa
\xigamone & = & \int_0^{\chi_s} \d\chi \int_1^{\infty} \frac{\d t}{t} 
\int_0^{2\pi} \frac{\d\phi}{\pi} \; (\wh\cos2\phi)^2 \Igamone \nonumber \\ 
& = & \frac{1}{2} \int_0^{\chi_s} \d\chi \wh^2 \pi \int_0^{\infty} 
\frac{\d k}{k} \; \frac{\Delta^2(k,z)}{k} \; W(k \De \theta_s)^2 .
\label{xigam1h1}
\eeqa
Here we have used $\tS_2=1$; see eq.(\ref{corrstar1}), which is still verified 
by our parameterization (\ref{zetadef}). Of course, eq.(\ref{xigam1h1}) is 
exact, within the small-angle approximation, and the last equality in 
eq.(\ref{xigam1h1}) shows that we recover the property $\lag \gamonehs^2 \rag 
= \lag \kaphs^2 \rag/2$. For the shear component, $\gam_2$, we obtain the same 
equations with $\cos2\phi$ replaced by $\sin2\phi$. This leaves the cumulants 
$\lag \gamtwohs^p \rag_c = \lag \gamonehs^p \rag_c$ unchanged and we get the 
same generating function. Therefore, we recover $\cP(\gam_1)= \cP(\gam_2)$. 
Finally, the PDF is given by the inverse Laplace transform (\ref{Pmu2}):
\beq
\cP(\gamonehs) = \inta \frac{\d y}{2\pi i \xigamone} \; 
e^{[\gamonehs y-\phigamone(y)]/\xigamone} .
\label{Pgam1h1}
\eeq
The generating function $\varphi(y)$ obtained in Section~\ref{Amplitude of 
the density correlations} shows a branch cut along the negative real axis 
for $y<y_s$; see eq.(\ref{ys}). Then, we see from eq.(\ref{phigam1}) that 
the generating function $\phigamone(y)$ for the shear component $\gam_1$ 
shows two symmetric branch cuts along the real axis, for $y<-\ysgamone$ and 
$y>\ysgamone$, with:
\beq
\ysgamone = \min_{z,t} \left| y_s \; \frac{\xigamone}{\Igamone \wh} \right| .
\label{ysgamone}
\eeq
Let us recall here that the generating function $\varphi(y;z)$ explicitly
depends on the redshift $z$ along the line of sight through the skewness
$S_3(z)$ of the density contrast at the typical wavenumber $k_s(z)$ probed 
at this redshift, see Section~\ref{Amplitude of the density correlations}.
Hence its singularity $y_s(z)$ also depends on the redshift $z$. Next,
the singularities of $\phigamone(y)$ at $\pm \ysgamone$ also
imply an exponential tail for $\cP(\gamonehs)$ at large $|\gamonehs|$:
\beq
|\gamonehs| \gg 1 : \hspace{0.3cm} \cP(\gamonehs) \sim 
e^{-\ysgamone |\gamonehs| /\xigamone} ,
\label{Pgam1hexp}
\eeq
which translates into similar exponential tails for the shear 
$\gam_{1\theta}=|\kappamin| \gamonehs$.

\begin{figure}
\protect\centerline{
\epsfysize = 3.5truein
%\epsfbox[27 75 477 564]
\epsfbox[25 147 588 715]
{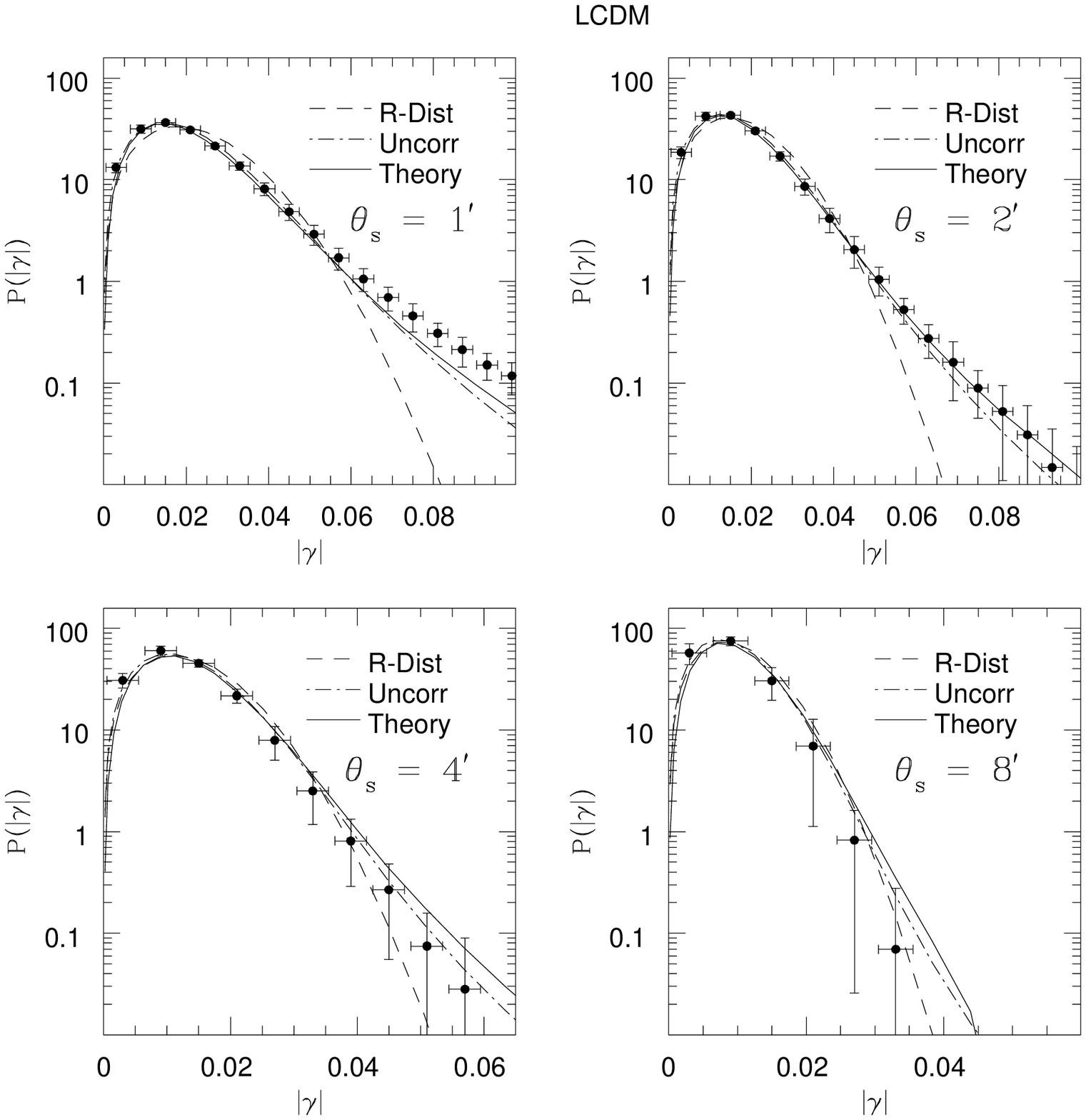} }
\caption{Probability distribution function, $\cP(|\gamma|)$, for the
LCDM cosmology, plotted as a function of $|\gamma|$. The smoothing
angle, $\theta_s$, varies from $1'$ (upper left panel) to $8'$ (lower
right panel). The data-points are the results from the numerical
simulations and the solid lines correspond to the analytical
predictions (\ref{Pmgam1}), (\ref{phimgam1}) and (\ref{Pgam12h1}) based on 
our stellar-model. The dotted-dashed curve assumes that 
the two shear components, $\gam_1$ and $\gam_2$, are uncorrelated but 
their PDF is given by our non-linear prediction; this corresponds to 
eqs.(\ref{phimgamuncorr1}) and (\ref{Pgam12uncorr1}). The dashed lines
are the predictions obtained for a Gaussian density field with the correct
non-linear variance (this yields a Rayleigh distribution for $|\gam|$).
The source redshift is fixed at unity in each case.}
\end{figure}

\subsection{PDF of the shear modulus, $|\gam|$}
\label{The shear modulus}

From the smoothed shear components, $\gamonehs$ and $\gamtwohs$, we may 
define the smoothed modulus, $\mgamhs$, as:
\beq
\mgamhs = | \gamonehs + i \gamtwohs | = \sqrt{\gamonehs^2 + \gamtwohs^2} .
\label{mgam1}
\eeq
Note that eq.(\ref{mgam1}) means that we take the modulus of the shear
{\it after} smoothing of the two components, $\gam_1$ and
$\gam_2$. Since smoothing and taking the modulus do not commute, the
quantity $\mgamhs$ defined in eq.(\ref{mgam1}) is {\it not} the
smoothed shear obtained by first taking the modulus
$|\gam|=|\gam_1+i\gam_2|$ at each point on the sky and second applying
a top-hat smoothing to this map. In this paper we shall only consider
the smoothed modulus $\mgamhs$ as defined in
eq.(\ref{mgam1}). Indeed, since taking the modulus is a non-linear
process one cannot directly apply the method used in this paper to
derive the PDF of the modulus $\gam$ defined by first taking the
modulus of the shear. By contrast, we can easily study the smoothed
modulus $\mgamhs$ defined in eq.(\ref{mgam1}) because it can be
directly obtained from the properties of the smoothed shear components,
$\gamonehs$ and $\gamtwohs$. Indeed, from eq.(\ref{mgam1}) we
obtain the relation:
\beq
\cP(\mgamhs) = \mgamhs \int_0^{2\pi} \d\alpha \; 
\cP(\mgamhs \cos\alpha,\mgamhs \sin\alpha) ,
\label{Pmgam1}
\eeq
where we have introduced the joint PDF, $\cP(\gamonehs,\gamtwohs)$, 
of the two shear 
components, $\gamonehs$ and $\gamtwohs$. Of course, following the method used 
in Section~\ref{PDF of the shear components} for the shear components we 
can obtain the joint PDF, $\cP(\gamonehs,\gamtwohs)$, from the cumulants, 
$\lag \gamonehs^{p_1} \gamtwohs^{p_2}\rag_c$. Thus, proceeding along the 
lines of Section~\ref{PDF of the shear components} the analog of 
eq.(\ref{gam1h7}) now reads:
\beqa
\lag \gamonehs^{p_1} \gamtwohs^{p_2} \rag_c & = & \int_0^{\chi_s} \d\chi
\; \tS_{p_1+p_2} \; \wh^{p_1+p_2} \int_1^{\infty} \frac{\d t}{t} 
\int_0^{2\pi} \frac{\d\phi}{\pi} \nonumber \\ & & \times \; (-\cos2\phi)^{p_1} 
(-\sin2\phi)^{p_2} \; \Igamone^{p_1+p_2-1} .
\label{mgamh1}
\eeqa
Then, the generating function, $\Phi_{\gamonehs,\gamtwohs}(y_1,y_2)$, defined 
as in eq.(\ref{Phi1}) by:
\beq
\Phi_{\gamonehs,\gamtwohs}(y_1,y_2) =  \sum_{p_1,p_2=0}^{\infty} 
\frac{(-1)^{p_1+p_2}}{p_1! \; p_2!} \; \lag \gamonehs^{p_1} \gamtwohs^{p_2} 
\rag_c \; y_1^{p_1} y_2^{p_2} 
\label{Phiy1y21}
\eeq
can be written from eq.(\ref{mgamh1}) as:
\beqa
\lefteqn{ \Phi_{\gamonehs,\gamtwohs}(y_1,y_2) = \int_0^{\chi_s} \d\chi 
\int_1^{\infty} \frac{\d t}{t} \int_0^{2\pi} \frac{\d\phi}{\pi} \; 
\frac{1}{\Igamone} } \nonumber \\ & & \times \sum_{p=1}^{\infty} \frac{1}{p!} 
\; \tS_p \; (\Igamone \wh)^p \left( y_1\cos2\phi+y_2\sin2\phi \right)^p .
\label{Phiy1y22}
\eeqa
Using the resummation (\ref{phideltaR}) we get:
\beqa
\lefteqn{ \Phi_{\gamonehs,\gamtwohs}(y_1,y_2) = - \int_0^{\chi_s} \d\chi 
\int_1^{\infty} \frac{\d t}{t} \int_0^{2\pi} \frac{\d\phi}{\pi} \; 
\frac{1}{\Igamone} } \nonumber \\ & & 
\times \; \varphi \left( - (y_1\cos2\phi+y_2\sin2\phi) \Igamone \wh 
; z \right) .
\label{Phiy1y23}
\eeqa
Next, from the cumulant generating function, $\Phi_{\gamonehs,\gamtwohs}$, 
we define the normalized generating function, $\varphi_{\gamonehs,\gamtwohs}$, 
by:
\beq
\varphi_{\gamonehs,\gamtwohs}(y_1,y_2) = - \xigamone 
\Phi_{\gamonehs,\gamtwohs}
\left( \frac{y_1}{\xigamone},\frac{y_2}{\xigamone} \right)
\eeq
and eq.(\ref{Phiy1y23}) yields:
\beqa
\lefteqn{ \varphi_{\gamonehs,\gamtwohs}(y_1,y_2) = \int_0^{\chi_s} \d\chi 
\int_1^{\infty} \frac{\d t}{t} \int_0^{2\pi} \frac{\d\phi}{\pi} \; 
\frac{\xigamone}{\Igamone} } \nonumber \\ 
& & \times \; \varphi \left( (y_1\cos2\phi+y_2\sin2\phi) 
\frac{\Igamone}{\xigamone} \wh ; z \right) .
\label{phimgam1}
\eeqa
Then, the joint PDF, $\cP(\gamonehs,\gamtwohs)$, is given by:
\beqa
\cP(\gamonehs,\gamtwohs) & = & \inta \frac{\d y_1 \d y_2}
{(2\pi i \xigamone)^2} \nonumber \\ 
& & \times \; e^{[\gamonehs y_1 + \gamtwohs y_2 
- \varphi_{\gamonehs,\gamtwohs}(y_1,y_2)]/\xigamone} .
\label{Pgam12h1}
\eeqa
It is interesting to compare our results (\ref{phimgam1}) and (\ref{Pgam12h1}) 
with the prediction we would obtain if we had assumed that $\gamonehs$ and 
$\gamtwohs$ were uncorrelated, but still described by the PDF derived in 
Section~\ref{PDF of the shear components}. Then, the normalized generating 
function, $\varphi_{\gamonehs,\gamtwohs}(y_1,y_2)$ would simply be given by:
\beq
\varphi_{\gamonehs,\gamtwohs}^{\rm uncorr}(y_1,y_2) = \phigamone(y_1) 
+ \phigamone(y_2) .
\label{phimgamuncorr1}
\eeq
Substituting this into eq.(\ref{Pgam12h1}), we obviously recover the property:
\beq
\cP^{\rm uncorr}(\gamonehs,\gamtwohs) = \cP(\gamonehs) \cP(\gamtwohs) .
\label{Pgam12uncorr1}
\eeq
Therefore, the fact that $\varphi_{\gamonehs,\gamtwohs}$ obtained in 
eq.(\ref{phimgam1}) cannot be written as the sum (\ref{phimgamuncorr1}) 
shows that within our stellar model of Section~\ref{Stellar-model} for 
the density field, both shear components, $\gamonehs$ and $\gamtwohs$, are 
correlated, which is actually quite natural. Moreover, the comparison 
of (\ref{Pgam12h1}) with (\ref{Pgam12uncorr1}) will allow us to study 
numerically the influence of these correlations onto the PDF of the 
smoothed modulus $\mgamhs$.

Finally, we can note that for a Gaussian density field we simply
have $\varphi(y)=-y^2/2$, see eqs.(\ref{xibp})-(\ref{phideltaR}). This also
corresponds to $\tS_2=1$ and $\tS_p=0$ for $p \geq 3$. Hence all our results
are exact for the case of a Gaussian density field, where we simply need to
use for the generating function: $\varphi^{\rm Gauss}(y)=-y^2/2$. 
Then, we recover from eq.(\ref{phigam1}) that the shear components 
$\gam_1$ and $\gam_2$ are also Gaussian random variables, since we 
find $\phigamone^{\rm Gauss}(y) \propto y^2$. This
was actually obvious from the definition of the shear as a linear integral
over the density field, see eq.(\ref{gamh1}). On the other hand, we can
see from eq.(\ref{phimgam1}) that in the Gaussian case we have
$\varphi_{\gamonehs,\gamtwohs}^{\rm Gauss}(y_1,y_2) = 
\phigamone^{\rm Gauss}(y_1) + \phigamone^{\rm Gauss}(y_2)$, since the 
cross-correlation term $y_1\cos2\phi \times y_2\sin2\phi$ vanishes after 
integration over $\phi$. Thus, we recover the fact that for a Gaussian density
field the two shear components $\gam_1$ and $\gam_2$ along the directions (1,2)
are uncorrelated. This is not surprising since for a Gaussian density
field the Fourier modes $\delta(\bk)$ are uncorrelated. This yields a 
Rayleigh distribution for the shear modulus $|\gamhs|$.

\begin{figure}
\protect\centerline{
\epsfysize = 3.5truein
%\epsfbox[27 75 477 564]
\epsfbox[25 147 588 715]
{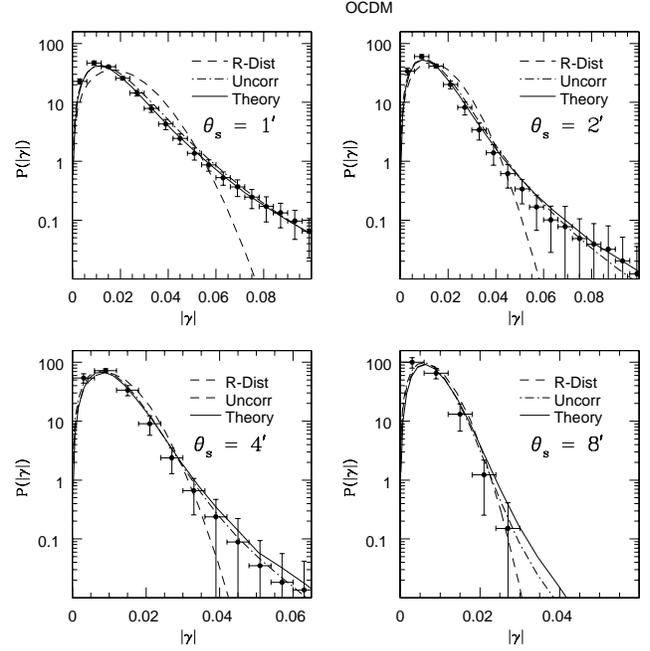} }
\caption{As for the previous Figure, but for the OCDM cosmology.}
\end{figure}

\section{Lensing statistics from numerical simulations}
\label{Numerical simulations}

We compare in the following Section our analytical results for the shear 
with weak
lensing statistics obtained in cosmological $N$-body simulations.
%REF
Couchman, Barber \& Thomas (1999) developed an algorithm for computing
the shear in three dimensions at locations within the simulation
volumes and we have applied this algorithm to the simulations of the
Hydra Consortium\footnote{(http://hydra.mcmaster.ca/hydra/index.html)}
produced using the `Hydra' $N$-body hydrodynamics code
%REF
(Couchman, Thomas \& Pearce, 1995).  

We used two simulations, a flat cosmology with a cosmological constant
and an open cosmology with zero cosmological constant. These will be
referred to as the LCDM and OCDM cosmologies, respectively. Both
contained dark matter particles only of mass $1.29 \times
10^{11}h^{-1}$ solar masses, where $h$ expresses the value of the
Hubble parameter in units of 100~km~s$^{-1}$~Mpc$^{-1}$. The number of
particles in each cosmology was $86^3$. We used a variable particle
softening in the code with a minimum value in box units of
$0.0007(1+z)$, where $z$ is the redshift of the simulation volume. The
simulation volumes had comoving side-dimensions of 100$h^{-1}$Mpc. To
avoid obvious structure correlations between adjacent boxes, each was
arbitrarily translated, rotated (by multiples of $90^{\circ}$) and
reflected about each coordinate axis, and in addition, each complete
run was performed 10 times in each cosmology. The cosmological and simulation
parameters of both the LCDM and OCDM simulations are given in Table 1.

\begin{table}
\begin{center}
\caption{Cosmological and simulation parameters
 characterizing the different models}
\label{tabsig2D}
\begin{tabular}{@{}lcccccc}
\hline
&$\Gamma$&$\Om$&$\Ol$&$\sigma_8$&$\theta_{\rm res}$&$\theta_{\rm survey}$\\
\hline
LCDM&0.25&0.3&0.7&1.22&$0'.34$&$2^{\circ}.6$ \\
OCDM&0.25&0.3&0.0&1.06&$0'.37$&$2^{\circ}.8$ \\
\hline
\end{tabular}
\end{center}
\end{table}

The general procedure for specifying the coordinates for the
lines-of-sight and the locations within the simulations for the
computations of the 3-d shear is as described by
%REF
Barber (2002). In the present work, a total of $455 \times 455$
lines of sight were used which completely filled the simulation box
immediately preceding the one at redshift 1 and allowed regular
sampling of the field of view. In addition, 300 regularly-spaced
evaluation locations for the shear along each line of sight were
specified in each simulation volume to adequately sample the varying
gravitational potential. In the LCDM cosmology, the full field of view
was $2^{\circ}.6 \times 2^{\circ}.6$ and in the OCDM cosmology,
$2^{\circ}.8 \times 2^{\circ}.8$.  The angular resolution in the
LCDM cosmology was $0'.34$, which equates to the minimum value of the
particle softening at the optimum redshift, $z = 0.36$, for lensing of
sources at a redshift of 1. In the case of the OCDM cosmology, the
angular resolution was $0'.37$. In both cosmologies, to allow for the
larger angular size of the minimum softening at low redshifts and also
for the range of particle softening scales above the minimum value, we
have not declared our results below angular scales of 1 arcminute.

To obtain the lensing statistics for sources at redshift $z_s = 1$, we
computed the required values assuming the sources were located at
the front face of the simulation volume whose redshift was closest to
1. Therefore in the LCDM cosmology, the sources were at $z_s = 0.99$
and in the OCDM cosmology they were at $z_s = 1.03.$

Once the 3-d shear had been computed at each location,
the multiple lens-plane theory, described clearly in
%REF
Schneider, Ehlers \& Falco (1992), was applied along the
lines of sight.  The procedure involved combining the intermediate
Jacobian matrices computed using the appropriate values of the angular
diameter distances in the relevant cosmology. Finally, the Jacobian
matrices appropriate at $z = 0$ for sources at $z_s = 1$ for every
line of sight were obtained.

From the Jacobian matrices in each of the 10 simulation runs for each
cosmology, we computed the shear components, $\gamma_1$ and
$\gamma_2$, and the full shear, $\gamma$. The values for the smoothed
components were computed on angular scales of $1'.0$, $2'.0$, $4'.0$
and $8'.0$, using a top-hat filter.

The computed values for the probability distributions from each of the
$N=10$ runs in each cosmology were averaged, and the errors on the
means of $1\sigma/\sqrt{N}$ determined.

\section{Comparison of Analytical predictions with results from 
numerical Simulations}

We compare here our analytical predictions for the smoothed shear
components and for the shear modulus with the results of the numerical
simulations. The quantites plotted are for various smoothing angles
from $1'$ to $8'$ in the two different cosmologies, LCDM and OCDM, and
in each case the source redshift is fixed at unity.

%Shear Components:  

Firstly we compare the results for the shear components, plotted in
Figure 2, for the LCDM cosmology, and Figure 3, for the OCDM
cosmology.  Since the PDFs for the components are even functions we only plot 
the PDF, $\cP(\gam_1)$, over $\gam_1 \geq 0$. The solid lines correspond 
to our 
analytical predictions, (\ref{phigam1}) and (\ref{Pgam1h1}), based on the 
stellar-model. For comparison, we also plot the Gaussian distribution 
which has the same variance (the dashed curves), so as to clearly show the
deviations from Gaussianity (this corresponds to a Gaussian underlying 
density field). We find that at small angular scales, 
$\theta_s \le 4'$, the results from the numerical simulations are very 
accurately described by our non-linear predictions where there are clear 
signs of departure from Gaussianity, especially in the tails of the PDF.
Of course, this merely reflects the non-linearity of the underlying mass 
distribution probed by these small angular windows. At larger smoothing 
angles, $\theta_s > 4'$, the stellar {\em Ansatz} seems to over-predict the
non-linearity induced by gravitational clustering and the results lie
between the fully non-linear predictions and the Gaussian
approximations. 
This behaviour may be due to the simple interpolation of eq.(\ref{S32})
for the transition between the linear and highly non-linear regimes, which
may not be sufficiently accurate. However, we did not try in this work to 
improve over the simple interpolation (\ref{S32}) because we expect the 
stellar model to be less accurate in the weakly non-linear regime.
Indeed, in the quasi-linear 
regime the angular dependence of the many-body correlations can be derived from
perturbation theory and one can see that for the three-point function it
already violates eqs.(\ref{xistar1})-(\ref{corrstar1}). On the other hand,
in the highly non-linear regime, the angular dependence of the many-body 
correlations gets shallower and eqs.(\ref{xistar1})-(\ref{corrstar1}) 
actually agree with numerical simulations for the three-point function 
(Scoccimarro \& Frieman 1999). Of course, at very large scales (which only
probe the linear regime) our results would become exact (within the small 
angle approximation) as we would recover the simple Gaussian.

\begin{figure}
\protect\centerline{
\epsfysize = 2.5truein
%\epsfbox[27 75 477 564]
\epsfbox[25 435 305 715]
{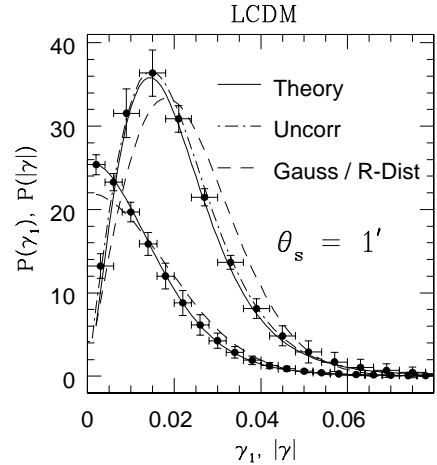} }
\caption{The PDFs, $\cP(\gam_1)$ and $\cP(|\gam|)$, on a linear-scale 
for the LCDM cosmology and smoothing angle $\theta_s=1'$. The source
redshift is again fixed at unity. The curves and the data-points are 
the same as in the upper left panels of Figures 2 and 4.}
\end{figure}

%Shear modulus:

The PDFs for the shear modulus are presented in Figure 4, for the LCDM
cosmology, and Figure 5, for the OCDM cosmology. The solid lines show our
analytical predictions (\ref{Pmgam1})-(\ref{phimgam1})-(\ref{Pgam12h1}) 
from the stellar-model. For comparison, we also plot 
(dotted-dashed curves) the results obtained by assuming that the two shear 
components are uncorrelated but their own PDFs are still given by our 
non-linear prediction; this corresponds to 
eqs.(\ref{phimgamuncorr1})-(\ref{Pgam12uncorr1}). Therefore, we see from 
Figures 4 and 5 that the correlation between both components, $\gam_1$ and 
$\gam_2$, only has a rather small influence on the PDF of the modulus, 
$\cP(|\gam|)$. For reference we have also shown the results for the 
Gaussian case (dashed curves), where the PDF for $|\gamma|$ 
is given by a Rayleigh distribution with a variance equal to that of the 
shear components (dashed curves). Comparison of our predictions with
the results of the numerical simulations shows that there is
a remarkable agreement in both cosmologies (as found for the shear
components), especially at the smaller smoothing scales. At the larger
scales, the agreement is less good for the same reasons as
described above. The sample variance arising from the various
realisations also makes the comparison more difficult; as the size of the
smoothing radius is increased, the number of patches with completely
independent information decreases, thereby increasing the sample
variance. In addition, although we made use of $10$ different
simulation runs (as described in Section 5), they were all based upon
the same $N$-body realisation in each cosmology, so the separate runs
may not contain completely independent information. Much larger
simulations would be required to quantify any departure from the
analytical predictions.

Finally, we display in Figure 6 our results over a linear-scale for $\cP$ 
rather than the logarithmic scale of the previous Figures, for the 
LCDM cosmology 
and the smoothing angle $\theta_s=1'$. While the previous Figures 
emphasized the
tails of the PDFs we can clearly see in Figure 6 the shapes of the PDFs near
their maxima. In agreement with the discussion above, we can check that
our predictions match the deviations from Gaussianity observed in the
numerical simulations. Thus, the PDF, $\cP(\gam_1)$, of the shear component,
$\gam_1$, is more sharply peaked around $\gam_1=0$ than a Gaussian.
Moreover, the location of the maximum of the PDF, $\cP(|\gam|)$, of the shear
modulus is accurately reproduced by our analytical model. These features
might be useful, in addition to the extended tails seen in the previous 
Figures, 
to measure in weak lensing surveys the departures from Gaussianity brought 
by the non-linear gravitational dynamics. This may also help to break the
degeneracies in the measurement of the cosmological parameters.

\section{Discussion}

%why our results are important

Extending earlier studies we have developed analytical techniques
based on a simple hierarchical {\em Ansatz}, which we called the
stellar model, to compute various statistical quantities associated
with cosmic shear. Thus, we have computed the full probability
distribution functions associated with the smoothed shear components,
$\cP(\gamma_i(\theta_s))$, and the modulus of cosmic shear,
$\cP(|\gamma(\theta_s)|)$. Here $|\gam(\theta_s)|$ is defined by first
smoothing both shear components and second taking their modulus. We
have also derived the joint-PDF,
$\cP(\gamma_1(\theta_s),\gamma_2(\theta_s))$. These results extend
similar studies applied to the convergence field. However, being a
spin-2 field, statistics from the shear are much richer than those
from the convergence field and, more importantly, can be directly used
for observational purposes where the non-trivial survey geometry can
pose a considerable problem for re-constructing convergence
$\kappa(\theta_s)$ maps.

The comparison of our analytical calculations with the results of the
numerical simulations shows that our simple method yields accurate
predictions at small angular scales, $\theta_s \le 4'$ (for a source
redshift $z_s=1$) which probe the non-linear regime for the underlying
density field. Therefore, we can use with confidence the simple
stellar model we introduced in this paper to study the shear
statistics and extract useful information from future surveys. In
particular, our analysis shows how the deviations from Gaussianity in
the density field built by the non-linear gravitational dynamics
translate into the PDF of the shear. This can also be seen in the
Figures where we drew Gaussian PDFs for comparison. Hence we may hope
to derive from shear maps some key properties of the structure of the
density field (e.g., the amplitude of the many-body correlations and
the shape of the PDF of the smoothed density contrast) which could
shed some light onto the non-linear regime of gravitational
clustering. In this respect, note that the separation of the shear,
$\gam$, into the factor $\kappamin$ and the ``normalized shear,''
$\gamh$, also provides a good separation between the dependence on the
cosmological parameters and the projection effects (e.g.,
$\Om,\Ol$ and $z_s$) versus the amplitude and the non-Gaussianities
of the fluctuations of the density field which characterize the collisionless 
gravitational dynamics itself. This is obvious for the convergence, $\kappa$, 
as recalled at the end of Section~\ref{PDF for the density field}, but
this feature also holds for the shear, $\gam$.

However, in the near future, where the observational noise may still
be important, one may first focus on the estimation of the
cosmological parameters. To this order, our study is complementary to
present analysis of the first few higher-order correlation functions
of the shear field (Zaldarriaga \& Scoccimarro, 2003, Schneider \&
Lombardi, 2003, Bernardeau, Mellier \& van Waerbeke, 2002, and
Takada \& Jain 2003) based on results derived for comological spin-2
fields such as CMB polarization, where shape dependence and scale
dependence of the shear is used to study non-Gaussianity induced by
the underlying mass distribution. In any case, a few measures of
non-Gaussianity (e.g., the skewness of the convergence, $\kappa$, see
Pen et al., 2003) are
useful to remove degeneracies between the various cosmological
parameters (e.g., $\Om$ vs $\sigma_8$) since a Gaussian is defined by
only one quantity (its variance), while we would like to measure
several quantities. Therefore, one may use our prediction for the PDF,
$\cP(\gam)$, to build some useful new observables.

% observations and cosmology

Before we can use our method to extract some information from 
weak gravitational lensing surveys, there are some points which need to be
addressed. In particular, we should take into account the effects of a finite 
width for the distribution of sources over redshift, as in actual surveys.
It would also be desirable to include realistic noise, such as
that arising from the intrinsic ellipticity distribution of galaxies,
and Poisson sampling due to the discreteness of the source
distribution. The finite size of the catalogue is also worthy of
consideration. These effects are beyond the scope of this paper, where we
have mainly investigated the accuracy of our approach to describe the physics
of weak gravitational lensing by large-scale structures. Having shown here
that our simple stellar model provides a useful basis, we plan to 
investigate these points in detail in future work.

% noise

To incorporate noise in our calculations we need to
convolve the pdf of shear which we have derived here with the noise pdf
which is generally assumed Gaussian. A complete analysis is beyond
the scope of the present paper. However, an order of magnitude estimate of the
signal to noise ratio for the kurtosis of the shear (lowest order moment) has 
been given by Takada \& Jain (2002). They conclude that for reasonable values
of survey parameters (rms intrinsic ellipticy $\sigma_\epsilon = 0.4$,
number density of source galaxies $n_{gal} = 30
{\rm ~arcmin}^{-2}$ and the survey area $\Omega_{survey} = {25 {\rm
~degree}}^2$)
it is possible to achieve a high signal to noise ratio for the small angular
scales which we have probed (although our cosmological models
differ slightly with the model they have considered but such estimates
are largely independent of the background cosmology). Since deviation from
non-Gaussianity in the case of shear pdf is dominated by kurtosis one would
expect roughly similar conclusions for the entire pdf too.
However a detailed analysis
in line with Munshi \& Coles (2002) will be presented in a future work.

% Born approximation

On the other hand, the present study already gives some information
about delicate points which are of theoretical and practical interest.
In particular, the underlying approximation in most analytical
calculations (including ours) is the Born approximation.  It neglects
higher order correction terms in the photon propagation equation. The
error introduced by such an approximation in the quasi-linear regime
has been studied by several authors (e.g. Schneider et al. 1998).
Although perturbative calculations tend to show that this error is
negligible for lower order cumulants, clearly such an analysis is not
possible in the highly non-linear regime. However, previous studies of
convergence statistics and our present results regarding the shear
PDF, which exhibit a good match between theoretical predictions and
the simulation results, indicate that such corrections are negligible
even in the highly non-linear regime.

%Different types of simulations

With regard to numerical simulations, previous work for the predicted
convergence statistics were made in comparison with ray-tracing
experiments. In the present case, our analytical results for the
shear, which are complimentary to the convergence results, are
compared successfully with the numerical computations based on
computing the full 3-d shear along every line of sight. The detailed
match of our various predictions not only verifies our analytical
calculations but also increases our confidence in the two different
numerical methods.

\section*{acknowledgements}
This work has been supported by PPARC and the numerical work carried
out with facilities provided by the University of Sussex. AJB was
supported in part by the Leverhulme Trust. The original code for the
3-d shear computations was written by Hugh Couchman of McMaster
University. There were many useful discussions with Antonio da
Silva. DM acknowledges the support from PPARC of grant RG28936.  DM
also acknowledges support from the University of Sussex during a visit
where part of the work was completed. It is a pleasure for DM to
acknowledge many fruitful discussions with members of Cambridge
Leverhulme Quantitative Cosmology Group, including Jerry Ostriker and
Alexandre Refregier.

\end{document}